\documentclass[aps,prd,floatfix,twocolumn,nofootinbib,superscriptaddress,showkeys, preprintnumbers]{revtex4-2}

\usepackage{amsmath,amssymb,bm,color,float,graphicx,mathrsfs}
\usepackage{hyperref}
\usepackage{Macro}
\hypersetup{colorlinks=true,linkcolor=blue,citecolor=blue,urlcolor=blue}

\usepackage{graphicx}

\begin{document}

\title{Alleviating CMB-BAO tension with a large cosmic birefringence angle during reionization induced by energy injection to axion-like particles}

\author{Salvatore Samuele Sirletti}\email{salvatore.sirletti@ipmu.jp}
\affiliation{Kavli Institute for the Physics and Mathematics of the Universe (WPI), UTIAS, The University of Tokyo, Kashiwa, 277-8583, Japan}

\author{Toshiya Namikawa}
\affiliation{Kavli Institute for the Physics and Mathematics of the Universe (WPI), UTIAS, The University of Tokyo, Kashiwa, 277-8583, Japan}

\author{Blake D. Sherwin}
\affiliation{Department of Applied Mathematics and Theoretical Physics, University of Cambridge,
Wilberforce Road, Cambridge CB3 0WA, United Kingdom}
\affiliation{Kavli Institute for Cosmology, University of Cambridge,Madingley Road, Cambridge CB3 0HA, United Kingdom}

\date{\today}

\begin{abstract}
The recent measurement of baryon acoustic oscillations by the Dark Energy Spectroscopic Instrument reveals a mild tension with observations of the cosmic microwave background (CMB) within the standard $\Lambda$CDM cosmological model. Another interesting hint beyond standard cosmology arises from recent analyses of \textit{Planck} polarization data which has been interpreted as cosmic birefringence. One possible solution to explain both observations is to introduce cosmic birefringence induced by axion-like particles (ALPs) that has a much larger rotation angle during the reionization epoch than that during the recombination: the large rotation suppresses the $EE$ reionization signal, allowing an elevated value of the CMB optical depth $\tau$ that alleviates the CMB-BAO tension. In this paper, we explore a scenario where the energy transfers from matter to ALPs before the reionization. Using \textit{Planck} low-$\ell$ $EE$, $BB$, and $EB$ data, we show that, for energy injection at $10 \lesssim z \lesssim 100$, there exist parameter regions where $\tau \simeq 0.08$--$0.10$ is compatible with the data.
\end{abstract}


\maketitle

\section{Introduction} \label{sec:intro}

Recent measurements of the cosmic microwave background (CMB) and baryon acoustic oscillations (BAO) exhibit a mild tension within spatially flat $\Lambda$ cold dark matter ($\Lambda$CDM) cosmology~\cite{DESI:2025:DR2,SPT-3G:2025}. In particular, cosmological analyses prefer a sum of neutrino masses below the minimum allowed by oscillation experiments~\cite{DESI:2025:neutrino}. When the effective neutrino mass is extended to negative values, the minimal mass is disfavored at approximately $3\sigma$, although the precise significance depends on the adopted \textit{Planck} likelihood~\cite{Craig:2024:neutrino,Green:2024:neutrino,Elbers:2024:neutrino,Lynch:2025:neutrino-tension,DESI:2025:neutrino,Herold:2024:mnu,Naredo-Tuero:2024:mnu}. 
One possible explanation is a shift in the CMB reionization optical depth, $\tau$~\cite{Giare:2023:tau,Loverde:2024:mnu,Sailer:2025:tau,Jhaveri:2025:tau}. If the \textit{Planck} low-$\ell$ polarization data are excluded, a larger value, such as $\tau=0.09$, substantially reduces the neutrino-mass tension and weakens the preference for dynamical dark energy, as well as the lensing and curvature anomalies~\cite{Giare:2023:tau,Sailer:2025:tau}. However, this value is inconsistent with the \textit{Planck} low-$\ell$ measurement at $3.8\sigma$~\cite{Sailer:2025:tau}. 

Another intriguing hint of beyond the $\Lambda$CDM model is cosmic birefringence, a rotation of the polarization plane of light during propagation (see Refs.~\cite{Alighieri:2015:review,Komatsu:2022:review} for reviews). Minami \& Komatsu \cite{Minami:2020:biref} first reanalyzed \textit{Planck} polarization data and reported a hint for isotropic birefringence at $2.4\,\sigma$ statistical significance. See also
Refs.~\cite{Gruppuso:2016nhj,Planck:2016soo,Ballardini:2025apf} for other analyses of the birefringence angle with Planck data. The analysis is improved by adding Wilkinson Microwave Anisotropy Probe polarization data, finding a rotation angle $\beta \simeq 0.34^{\circ}$ at a $3.6\,\sigma$ preference~\cite{Eskilt:2022:biref-const}. The latest constraint is obtained by a joint analysis of the \textit{Planck} and ACT data that yields a birefringence angle of $(0.277\pm0.057)^\circ$, which is non-zero with a statistical significance of $4.8\,\sigma$ \cite{Eskilt:2026}. 
Because cosmic birefringence violates parity, it provides a sensitive probe of new physics beyond both $\Lambda$CDM and the Standard Model~\cite{Nakai:2023}.
Cosmic birefringence can be generated by a pseudoscalar field $\phi$, such as an axion-like particle (ALP), coupled to electromagnetism through a Chern--Simons interaction,
\begin{equation}
    \mathcal{L}\supset -\frac{1}{4}g\phi F_{\mu\nu}\tilde{F}^{\mu\nu},
\end{equation}
where $g$ is the coupling constant, $F_{\mu\nu}$ is the electromagnetic field-strength tensor, and $\tilde{F}^{\mu\nu}$ is its dual. This mechanism has been studied in models of axion dark energy~\cite{Carroll:1998:DE,Liu:2006:biref-time-evolve,Panda:2010,Fujita:2020:forecast,Fujita:2020:biref,Choi:2021aze,Obata:2021,Gasparotto:2022uqo,Galaverni:2023, Lin:2025gne}, early dark energy~\cite{Fujita:2020:biref,Murai:2022:EDE,Eskilt:2023:EDE,Kochappan:2024:biref}, and axion dark matter~\cite{Finelli:2009,Sigl:2018:biref-sup,Liu:2016:AxionDM,Fedderke:2019:biref,Zhang:2024dmi,Namikawa:2025:Planck}. Other proposed sources include topological defects~\cite{Takahashi:2020tqv,Kitajima:2022jzz,Jain:2022jrp,Gonzalez:2022mcx,Lee:2025:biref-DW} and quantum-gravity effects~\cite{Myers:2003fd,Balaji:2003sw,Arvanitaki:2009fg}.

One possible scenario to explain the above two anomalies is to introduce a large rotation of cosmic birefringence during the reionization epoch \cite{Namikawa:2025:PRL}. If cosmic birefringence is induced by a single ALP, the ambiguity of the phase of the observed cosmic birefringence \cite{Naokawa:2024xhn} could lead to a large rotation angle during the reionization while keeping the recombination birefringence signal unchanged. 
Alternatively, we can also consider a scenario originally proposed in Graham et al. (2025) \cite{Graham:2025} where part of the energy of matter transfers to ALPs. 

In this paper, we develop the scenario of Ref.~\cite{Graham:2025} to quantitatively constrain that scenario. We use \textit{Planck} data to search for parameter regions that reproduce the observed data and to show whether this model explains the recent CMB-BAO tension.

This paper is organized as follows. In Sec.~\ref{sec:theory}, we review cosmic birefringence induced by ALPs and introduce the energy injection model. In Sec.~\ref{sec:results}, we present constraints on cosmic-birefringence scenarios using data from \textit{Planck}. Finally, Sec.~\ref{sec:discussion} is devoted to a summary of our results and conclusions.

\section{Cosmic Birefringence and Energy Injection Model} \label{sec:theory}

\subsection{CMB angular power spectra}

Let us first introduce the CMB angular power spectra induced by cosmic birefringence from ALPs~\cite{Liu:2006:biref-time-evolve,Finelli:2009,Gubitosi:2014:biref-time,Lee:2013:biref,Sherwin:2021:biref,Nakatsuka:2022,Murai:2022:EDE,Naokawa:2023,Murai:2024yul, Arcari:2024nhw}. 
When cosmic birefringence is sourced by an ALP field, the rotation angle becomes time-dependent. The net rotation angle for a photon observed today and emitted at conformal time $\eta$ is given by \cite{Carroll:1989:rot,Carroll&Field:1991,Harari:1992}
\al{
    \beta(\eta) 
    &= \frac{g}{2}[\phi(\eta_0)-\phi(\eta)] 
    = \frac{g\phi_{\rm ini}}{2}\frac{\phi(\eta_0)-\phi(\eta)}{\phi_{\rm ini}}
    \,, \label{Eq:beta}
}
where $\eta_0$ is the present conformal time and $\phi_{\rm ini}$ is the initial value of $\phi$. Throughout this paper, the overall amplitude $g\phi_{\rm ini}/2$ is fixed so that the present-day rotation angle of the CMB photons emitted at recombination is $\beta_0 \simeq 0.34^\circ$~\cite{Eskilt:2022:biref-const}. Since $\beta(\eta)$ scales linearly with $g\phi_{\rm ini}/2$, varying $\beta_0$ within its uncertainty would only shift the allowed values of the energy injection, without changing our results qualitatively.
To compute the impact of this time-dependent rotation on the CMB polarization, we solve the Boltzmann equation for the polarized photon distribution~\cite{Liu:2006:biref-time-evolve,Finelli:2009,Gubitosi:2014:biref-time,Lee:2013:biref,Kamionkowski&Caldwell:2026}:
\begin{align}
    &_{\pm2}\Delta'_P + \iu q\nu~_{\pm2}\Delta_P 
    \notag \\
    &\qquad = a n_{\rm e}\sigma_T
        \left[
            -~_{\pm2}\Delta_P + \sqrt{\frac{6\pi}{5}}~_{\pm2}Y_{20}(\nu)\Pi(\eta,q)
        \right]
    \notag \\
    &\qquad\qquad \pm \iu g\phi'~_{\pm2}\Delta_P
    \,, \label{Eq:Boltzmann}
\end{align}
where $_{\pm2}\Delta_P$ are the Fourier modes of the linear polarization $Q \pm \iu U$ and are functions of conformal time $\eta$, wavevector magnitude $q$, and angle cosine between line-of-sight direction and wavevector $\nu$. The other symbols are: $a$ (scale factor), $n_{\rm e}$ (electron number density), $\sigma_{\rm T}$ (Thomson scattering cross-section), $\Pi$ (polarization source term)~\cite{Zaldarriaga:1996xe}, and ${}_{\pm2} Y_{20}$ (spin-two spherical harmonics). Primes denote derivatives with respect to conformal time. 

To derive the angular power spectra, $_{\pm2}\Delta_P$ is expanded as~\cite{Zaldarriaga:1996xe}:
\begin{align}
    _{\pm2}\Delta_P(\eta_0,q,\nu) 
    &\equiv -\sum_\l \iu^{-\l}\sqrt{4\pi(2\l+1)} 
    \notag \notag \\
    &\quad\times [\Delta_{E,\l}(q)\pm \iu\Delta_{B,\l}(q)] {}_{\pm2}Y_{\l0}(\nu) 
    \,, 
\end{align}
where the formal solution for the $E$- and $B$-mode sources is given as a line-of-sight integral \cite{Liu:2006:biref-time-evolve}: 
\begin{align} \label{eq:Deltas}
    &\Delta_{E,\l}(q)\pm \iu\Delta_{B,\l}(q) = \frac{3}{4}\sqrt{\frac{(\l+2)!}{(\l-2)!}}
    \notag \\
    &\qquad\times \INT{}{\eta}{}{0}{\eta_0}an_{\rm e}\sigma_T \E^{-\tau(\eta)}\Pi \frac{j_l(x)}{x^2}\E^{\pm 2\iu \beta(\eta)}
    \,. 
\end{align}
Here, $j_\l(x)$ is the spherical Bessel function, $x=q(\eta_0-\eta)$, and $\tau(\eta)=\int_{\eta}^{\eta_0}{\rm d}\eta_1 an_{\rm e}\sigma_T$. Note that the $E$-mode generated at time $\eta$ is multiplied by $\cos[2\beta(\eta)]$. Thus, a large variation of $\beta(\eta)$ during reionization leads to cancellations of signals among different epochs, resulting in a suppression of the reionization bump signal in the $EE$ power spectrum and mimicking a lower $\tau$~\cite{Namikawa:2025:PRL}. The CMB angular power spectrum is then calculated as
\begin{equation}
    C_\l^{XY} = 4\pi
    \Int{}{(\ln q)}{} \mathcal{P}_s(q)\Delta_{X,\l}(q)\Delta_{Y,\l}(q)
    \,, \label{Eq:ClXY}
\end{equation}
where $X, Y \in {E, B}$ and $\mathcal{P}_s(q)$ is the dimensionless primordial scalar curvature power spectrum. We systematically compute all CMB polarization auto- and cross-power spectra in Eq.~\eqref{Eq:ClXY} by appropriately modifying the polarization transfer function, $\Delta_{X,\ell}$, within the {\tt CLASS} code \cite{CLASS, Blas:2011rf}. 

\subsection{Energy injection from CDM to ALPs}\label{subsec:En_inj}

We now incorporate a large rotation angle realized by energy injection to ALPs from CDM. 
We consider the following phenomenological model approach often used for CDM / dark energy interactions ~\cite{Gavela:2009cy, Salvatelli:2013wra, DiValentino:2017iww,Wang:2016lxa, Graham:2025, DiValentino:2019exe}; we modify the continuity equation that transfers part of the DM energy density to ALPs: 
\al{
    &\bar{\rho}'_{\rm dm} + 3\mathcal{H}\bar{\rho}_{\rm dm} = -Q
    \,, \label{eq:densityeqCDM}
    \\ 
    &\bar{\rho}'_{\phi} + 3\mathcal{H}(\bar{\rho}_{\phi}+\bar{p}_{\phi}) = Q
    \,, \label{eq:densityeqALP}
}
where $\bar{\rho}_{\rm dm}$ is the energy density of CDM, $\bar{\rho}_{\phi}$ is the energy density of ALPs, $\bar{p}_\phi$ is the pressure of ALPs, $\mC{H}=a'/a$, $Q$ is the rate of energy transfer. 

We rewrite the continuity equation in terms of the pseudo-scalar field as
\al{
    \frac{\phi'}{a^2}(\phi''+2{\cal H}\phi'+a^2V_{,\phi}) = Q
    \,, \label{Eq:EoM:phi-Q}
}
where $V_{,\phi}(\phi)$ is the partial derivative of the potential $V(\phi)$ with respect to $\phi$. In the following, we assume a cosine potential $V(\phi)=m_\phi^2f_\phi^2[1-\cos(\phi/f_\phi)]$ where $f_\phi$ is the decay constant. 
We assume that ALP is a spectator field and that the energy density of ALPs is negligible in the background evolution. Thus, even a small amount of energy transfer from matter can significantly enhance the energy density of ALPs, and the amplitude of $\phi$. 
Note that in the absence of the energy transfer, we obtain the usual Klein-Gordon equation: 
\begin{equation}
    \phi'' + 2\mC{H}\phi' + a^2V_{,\phi} = 0
    \,. \label{Eq:phi-EoM}
\end{equation}
We assume the following energy transfer rate (e.g., \cite{DiValentino:2017iww, Wang:2016lxa})
\begin{equation}\label{eq:Q}
     Q = \mathcal{H}\xi\bar{\rho}_{\rm dm}
     \,.
\end{equation}
Instead of using a specific functional form for $\xi$, we constrain $\xi$ in a model independent way; we assume that $\xi$ is a constant during a specific short period starting at the redshift of $z=z_{\rm start}$ and ending at $z_{\rm end}$. 
Substituting the above equation into Eq.~\eqref{Eq:EoM:phi-Q} yields 
\begin{equation}\label{eq:EoM_phi}
        \phi'' + 2\mathcal{H}\phi' + a^2V_{,\phi} = \frac{a^2\mathcal{H}\xi\bar{\rho}_{\rm dm}}{\phi'}
    \,, 
\end{equation}
when $\phi'\not=0$. In our calculation, $\phi'\not=0$ is always satisfied during the energy injection period.

To realize a large rotation angle, the ALP field is initially located near the bottom of the potential. We specifically assume the initial condition that satisfies 
\al{
    &\theta_{\rm ini}\equiv \frac{\phi_{\rm ini}}{f_\phi} = 0.01\ll 1
    \,, \\ 
    &\phi'_{\rm ini} = 0
    \,. 
}
This choice, together with $\beta_0 \simeq 0.34^\circ$, implies $g f_\phi \simeq 1$, i.e., an anomaly coefficient $C = 2\pi g f_\phi/\alpha \sim 10^3$ for $g = \alpha C/(2\pi f_\phi)$. Such large values have been
considered in previous models of cosmic birefringence~\cite{Choi:2021aze}. In this case, the initial ALP energy density is $\bar\rho_{\phi,\rm ini} = V(\phi_{\rm ini})\simeq m_\phi^2\phi_{\rm ini}^2/2$, where the last equality holds at the bottom of the potential well. Considering that $\bar{\rho}_{\rm dm} = \bar\rho_{\rm dm,0} a^{-3}$, and denoting $x\equiv\phi/\phi_{\rm ini}$, the evolution equation is rewritten as
\begin{equation}
    \label{Eq:x-EoM:inject}
    x'' + 2{\cal H}x' + a^2V_{,x}(x) = \frac{{\mathcal H}}{ax'}\frac{m_\phi^2}{2}\xi\frac{\bar\rho_{\rm dm,0}}{\bar\rho_{\phi,\rm ini}}
    \,, 
\end{equation}
where $\bar{\rho}_{\rm dm,0}$ is the present energy density of DM and $V(x) = m_\phi^2/\theta_{\rm ini}^2 \left[1 - \cos(\theta_{\rm ini}\, x)\right]$. 

It is convenient to define
\al{
    F &\equiv \frac{\bar\rho_{\phi,\rm ini}}{\bar\rho_{{\rm dm},0}}
    \,, \\
    \tilde\xi &\equiv \frac{\xi}{F} = \xi\,\frac{\bar\rho_{{\rm dm},0}}{\bar\rho_{\phi,\rm ini}} 
    \,.
}
For fixed $\theta_{\rm ini}=0.01$, Eq.~\eqref{Eq:x-EoM:inject} depends on $\xi$ and $F$ only through $\tilde\xi$: the evolution of $x=\phi/\phi_{\rm ini}$, the birefringence angle, and the CMB spectra are all functions of $\tilde\xi$ alone. $F$ (and also $\xi$) is instead a free parameter that the CMB spectra cannot determine. $F$ only rescales the physical injection rate $\xi=\tilde\xi F$ and the initial ALP energy, without affecting $\tilde\xi$. 

\begin{figure}[t]
\centering
\includegraphics[width=\columnwidth]{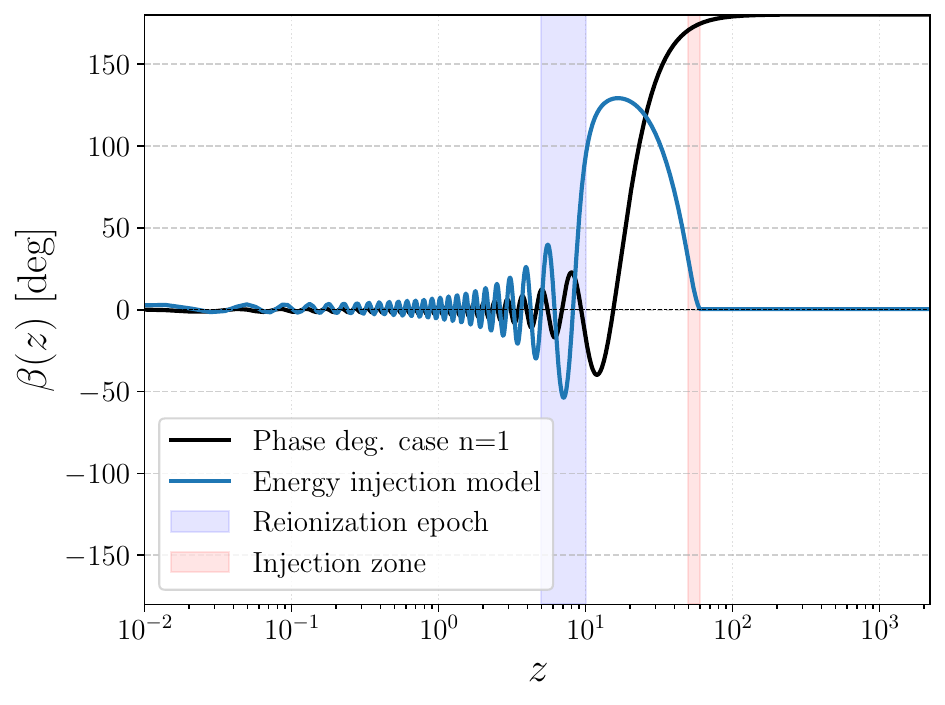}
\caption{Time evolution of the cosmic birefringence angle $\beta(z)$ for
$\log_{10}(m_\phi/\mathrm{eV}) = -30.6$ and an energy-injection of $\tilde\xi = 22$ (blue curve), with injection occurring in the redshift interval $50 \lesssim z \lesssim 60$ (red band). The blue band marks the reionization epoch as reference. The rotation angle is null before the injection takes place; once the injection begins, $\phi$ is kicked out of its potential minimum and starts to oscillate, producing the large swing in $\beta(z)$. The oscillations are subsequently damped by Hubble friction, and $\beta(z)$ asymptotically approaches the present-day birefringence angle $\beta_0 \simeq 0.34^\circ$. The $n\pi$ phase model (black curve) used in \cite{Namikawa:2025:PRL} is shown for comparison.}
\label{fig:beta_injection}
\end{figure}

Figure~\ref{fig:beta_injection} shows an example of the evolution of the ALP field. We choose $\tilde\xi\sim \mathcal{O}(10)$ that leads to a large birefringence angle. In the following, we focus on the rescaled energy injection rate $\tilde{\xi}$ instead of $\xi$. 

\begin{figure}[t]
\centering
\includegraphics[width=\columnwidth]{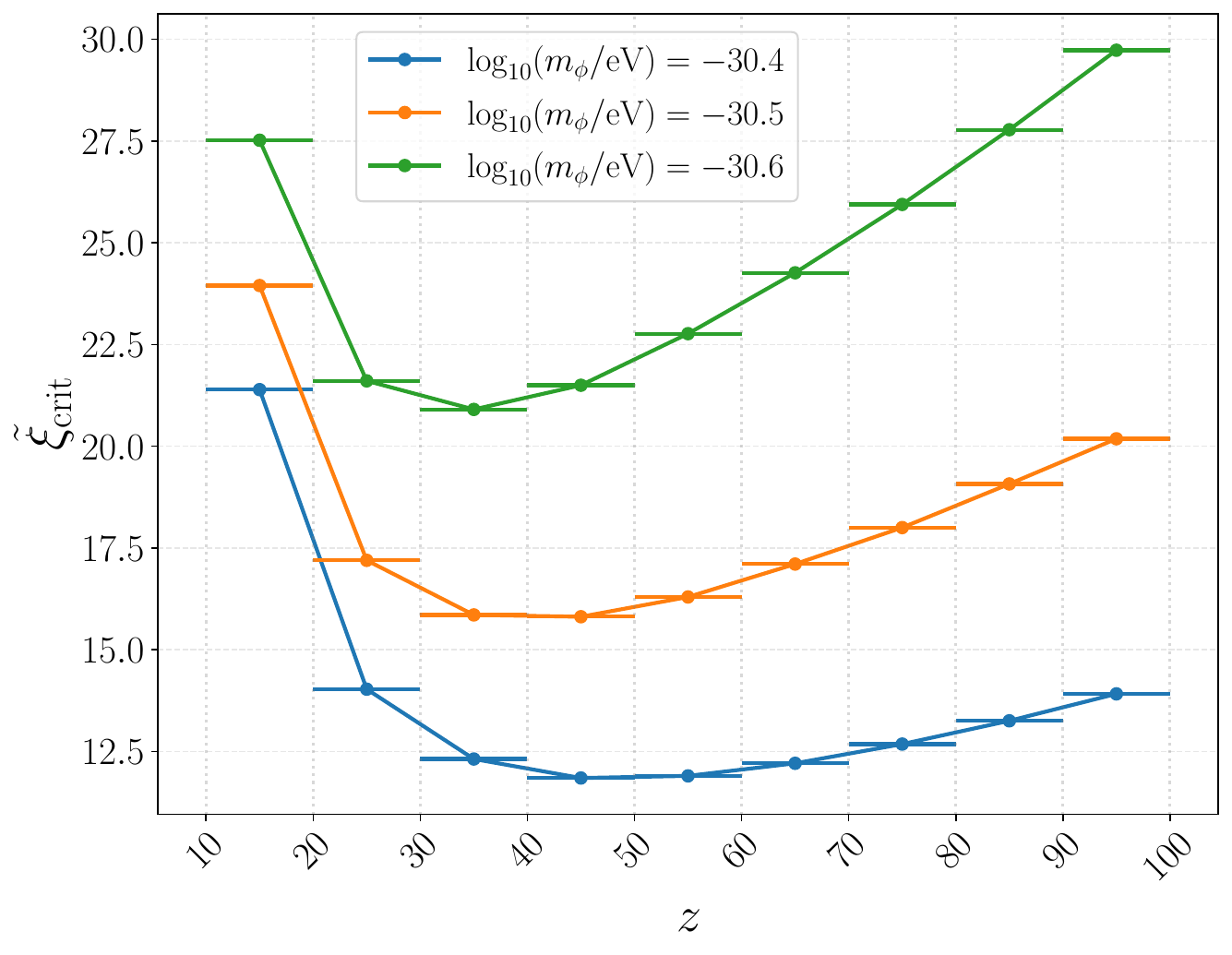}
\caption{Critical energy-injection efficiency $\tilde\xi_{\rm crit}$ as a function of the injection redshift window, for $\log_{10}(m_\phi/\mathrm{eV}) = -30.4, -30.5, -30.6$. The non-monotonic trend, with a minimum around $z \sim 30$--$50$ that shifts to lower redshift for lighter masses, reflects the efficient coupling between the injection and the epoch $H(z) \sim m_\phi$, when the field starts to be released from Hubble friction.}
\label{fig:xicrit}
\end{figure}

We note that $\tilde{\xi}$ has an upper bound to realize a large birefringence angle. 
To see this, let us consider the case if too much energy is injected into ALPs by increasing $\tilde\xi$, and $\phi$ crosses over the peak of the cosine potential, $\phi = k\pi f_\phi$, $k \in \mathbb{Z}$. In this situation, the ALP no longer oscillates around $\phi=0$ and instead settles into oscillations around a different minimum, $\phi_{\rm min}$. Then, the resulting cosmic birefringence angle for the CMB photons emitted at recombination becomes approximately $g\phi_{\rm min}/2\simeq 0.34^\circ$. Since the field value varies within $\sim \phi_{\rm min}$, we cannot realize a large rotation angle. 
Thus, there is a critical value of the energy transfer efficiency, $\tilde\xi = \tilde\xi_{\rm crit}$, beyond which this kick-out occurs. 
Because the driving term is time-dependent and the injection acts only over a finite redshift interval, $\tilde\xi_{\rm crit}$ cannot be obtained in closed form and is instead determined numerically for each ALP mass and each injection window. Figure~\ref{fig:xicrit} shows $\tilde\xi_{\rm crit}$ as a function of the injection redshift window for $\log_{10}(m_\phi/\mathrm{eV}) = -30.4, -30.5, -30.6$. For the three representative masses shown, $\tilde\xi_{\rm crit}(z)$ is non-monotonic, exhibiting a minimum around $z \sim 30$--$50$. This minimum marks the redshift bin where energy injection couples most efficiently to the field, corresponding to the epoch when the field starts to be released from Hubble friction, $H(z) \sim m_\phi$, shortly before the onset of coherent oscillations at $m_\phi \simeq 3H(z_{\rm osc})$: injecting energy at this epoch requires the smallest $\tilde\xi$ to kick the field over the potential barrier, while injection well before (field still frozen by Hubble friction) or well after (field already oscillating and decaying) is less efficient and requires larger $\tilde\xi$. Consistently, the minimum shifts to lower redshift for the lighter ALP, tracking the mass dependence of the epoch $H(z) \sim m_\phi$.

\subsection{Constraints on $\xi$ and $F$}\label{sec:constraints_xiF}

As discussed in Sec.~\ref{subsec:En_inj}, the CMB spectra depend on the energy injection only through $\tilde\xi=\xi/F$, where $F=\bar\rho_{\phi,\rm ini}/\bar\rho_{{\rm dm},0}$ is a free parameter. Consequently, $\xi$ and $F$ are degenerate and cannot be independently constrained by the CMB spectra alone. However, the physical energy exchanged between CDM and ALPs depends on both $\xi=\tilde\xi F$ and $F$. Since the energy densities of CDM and ALPs are constrained by CMB and large-scale structure observations, these constraints impose additional restrictions on the allowed values of $\xi$ and $F$. We consider the following two conditions:
\begin{enumerate}
    \item \textit{CDM depletion}: The energy transferred to the ALP is extracted from CDM. Since the CDM energy density is tightly constrained to be close to that in the best-fit $\Lambda$CDM cosmology, the fraction of CDM depleted must be smaller than the observational uncertainty in the CDM energy density. As the CDM energy density is proportional to the physical density parameter $\Omega_{\rm dm}h^2$, we conservatively require
    \begin{equation}\label{eq:deplationlimit}
        1- \frac{\bar\rho_{\rm dm}}{\bar\rho_{\rm dm}^{\Lambda{\rm CDM}}}\bigg|_{a=1} < \sigma(\Omega_{\rm dm}h^2) \simeq 10^{-2}
        \,.
    \end{equation}
    Here, the $1\,\sigma$ uncertainty is taken from Ref.~\cite{PR3:main}.
    \item \textit{Spectator condition}: The ALP energy density must remain negligible compared to that of CDM; otherwise, it could significantly modify the background expansion history, potentially conflicting with large-scale structure observations. Following Ref.~\cite{Fujita:2020:biref}, we impose $\Omega_\phi h^2<0.006$, where $\Omega_\phi$ denotes the present-day ALP density parameter. This condition can be more conservatively expressed as
    \begin{equation}\label{eq:spectatorlimit}
        \frac{\bar\rho_{\phi,0}}{\bar\rho_{\rm dm,0}} = \frac{\Omega_\phi h^2}{\Omega_{\rm dm}h^2} \lesssim 10^{-2}
        \,,
    \end{equation}
    where we have used $\Omega_{\rm dm}h^2\simeq 0.1$ \cite{PR3:main}.
\end{enumerate}
These two conditions do not alter the predicted CMB spectra at fixed $\tilde\xi$, but they restrict the values of $\xi$ and $F$ for which a given $\tilde\xi$ is physically viable.

Satisfying these conditions also justifies the approximations adopted in our numerical calculations: the background CDM energy density is fixed to its best-fit $\Lambda$CDM value, and the ALP is treated as a spectator field whose contribution to the background expansion is negligible.

In the following, we derive constraints on $\xi$ and $F$ from these two conditions.
\\

\textit{CDM depletion.}
Integrating Eq.~\eqref{eq:densityeqCDM} with $Q$ of Eq.~\eqref{eq:Q} gives $\bar\rho_{\rm dm}\propto a^{-(3+\xi)}$ for $z_{\rm end}\le z\le z_{\rm start}$ and $\bar\rho_{\rm dm}\propto a^{-3}$ otherwise. The ratio to the $\Lambda$CDM density after the injection is therefore
\begin{equation}\label{eq:depletion_eq}
    R(\xi)\equiv\frac{\bar\rho_{\rm dm}}{\bar\rho_{\rm dm}^{\Lambda{\rm CDM}}}=\left(\frac{1+z_{\rm start}}{1+z_{\rm end}}\right)^{-\xi} 
    \,,
\end{equation}
and Eq.~\eqref{eq:deplationlimit} becomes a limit on $\xi$ only,
\begin{equation}\label{eq:xi_dep}
    \xi < \xi_{\rm dep} = -\frac{\ln(0.99)}{\ln \left(\frac{1+z_{\rm start}}{1+z_{\rm end}} \right)} \,,
\end{equation}
which depends only on the injection window and is independent of $m_\phi$ and $F$.

\textit{Spectator condition.}
The ALP energy density has two contributions: the initial energy density $\bar\rho_{\phi,\rm ini}$, frozen in the potential until the field starts to oscillate, and the energy injected by the source term $Q$. The ratio $\bar\rho_\phi/\bar\rho_{\rm dm}$ grows while the field is frozen, since $\bar\rho_{\rm dm}$ dilutes as $a^{-3}$; once the field starts to oscillate, $\bar\rho_\phi$ also dilutes as $a^{-3}$ and the ratio settles to a plateau. We therefore evaluate Eq.~\eqref{eq:spectatorlimit} today, $a=1$, where the plateau value is reached and which is representative of the maximum of the ratio over the whole evolution. Using the results in Appendix~\ref{app:phi_thermodynamics}, we find
\begin{equation}\label{eq:ratio_a1}
  \frac{\bar\rho_{\phi,0}}{\bar\rho_{\rm dm,0}}=F\,\frac{1}{R(\xi)}\left[a^3\frac{\bar{\rho}_\phi(a)}{\bar{\rho}_{\phi,\rm ini}}\right]_{a=1}\,.
\end{equation}
This ratio is not simply proportional to $F$ at fixed $\xi$: Eq.~\eqref{Eq:x-EoM:inject} depends on $F$ and $\xi$ only through $\tilde\xi$, so that $\bar\rho_\phi(a)/\bar\rho_{\phi,\rm ini}$ is a function of $\tilde\xi$. Defining the transfer function
\begin{equation}\label{eq:mathcalG}
\mathcal{G}(\xi,F)\equiv\left[a^3\frac{\bar{\rho}_\phi(a)}{\bar{\rho}_{\phi,\rm ini}}\right]_{a=1}=\mathcal{G}(\tilde\xi)\,,
\end{equation}
which depends on $\xi$ and $F$ only through their ratio $\tilde\xi=\xi/F$, Eq.~\eqref{eq:ratio_a1} reads

\begin{equation}
  \frac{\bar\rho_{\phi,0}}{\bar\rho_{\rm dm,0}}=F\,\frac{\mathcal{G}(\tilde\xi)}{R(\xi)}.
\end{equation}

The function $\mathcal{G}(\tilde\xi)$ includes non-linear effects of the injection, such as the kinetic phase that follows a strong kick and the delay in the onset of oscillations, which analytical approximations cannot reproduce (Appendix~\ref{app:phi_thermodynamics}). We therefore compute it numerically, solving Eq.~\eqref{Eq:x-EoM:inject} once for each $\tilde\xi$ on a grid $0\le\tilde\xi\le\tilde\xi_{\rm crit}$, and then map it onto the $(\xi,F)$ plane using $\tilde\xi=\xi/F$. The allowed region is the intersection of: 
\begin{enumerate}
    \item the kick-out limit, $\tilde\xi\le\tilde\xi_{\rm crit}(m_\phi)$;
    \item the spectator limit, $F\,\mathcal{G}(\tilde\xi)/R(\xi)\le10^{-2}$;
    \item the depletion limit $\xi<\xi_{\rm dep}$.
\end{enumerate}

\begin{figure}[t]
\centering
\includegraphics[width=\columnwidth]{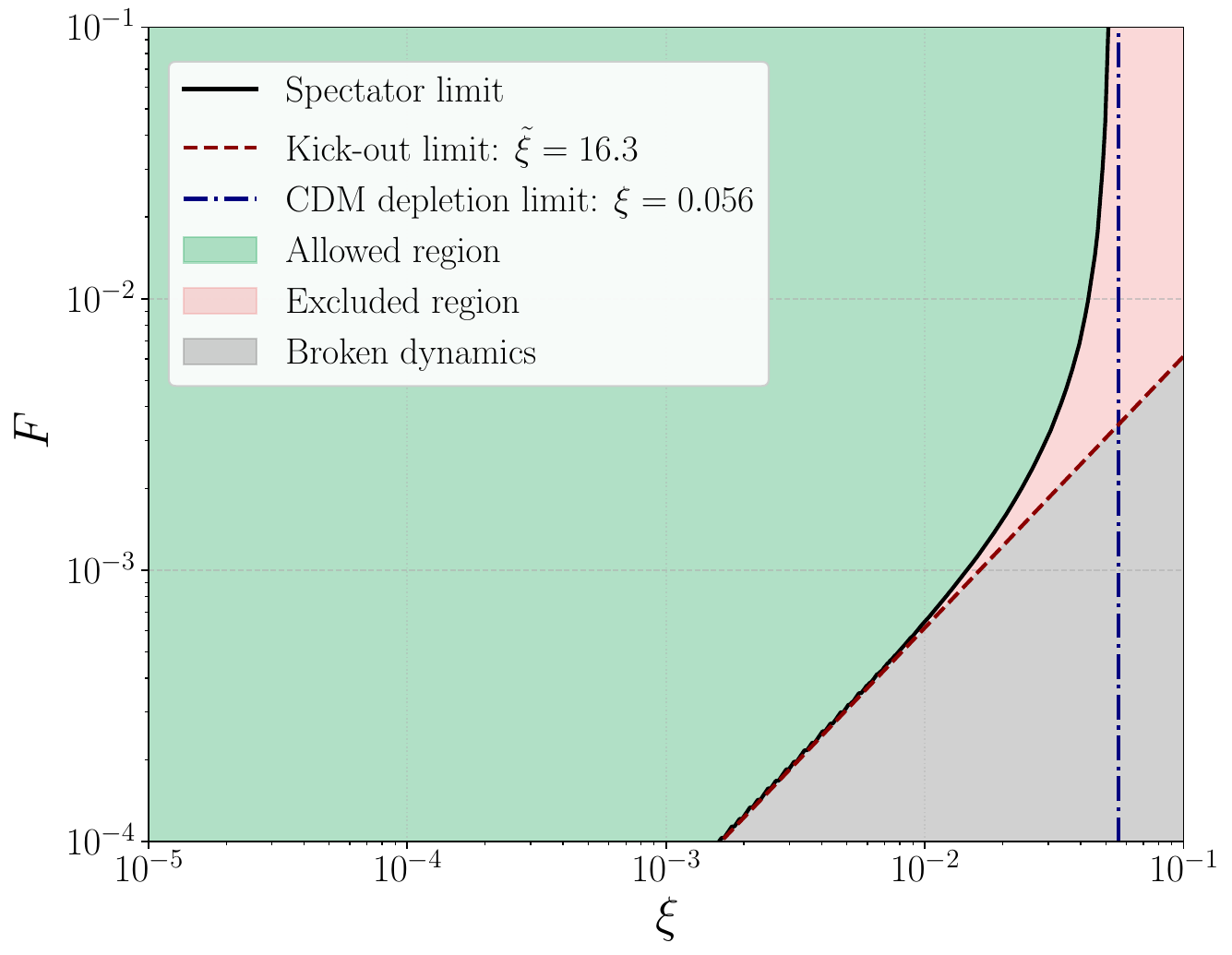}
\caption{Allowed region in the $(\xi,F)$ plane for $\log_{10}(m_\phi/{\rm eV})=-30.5$ and $50\le z\le60$. Black: spectator limit, $\bar\rho_\phi/\bar\rho_{\rm dm}=10^{-2}$; dashed red: kick-out limit $\xi=\tilde\xi_{\rm crit}F$ with $\tilde\xi_{\rm crit}=16.3$ (gray: broken dynamics); dash-dotted blue: CDM depletion limit $\xi=\xi_{\rm dep}$.}
\label{fig:exclusion_plot_logm-30.5}
\end{figure}

Figure~\ref{fig:exclusion_plot_logm-30.5} shows the result for $\log_{10}(m_\phi/\mathrm{eV})=-30.5$.
Since the CMB spectra depend only on $\tilde\xi$, the choice of $F$ only moves the physical value of $\xi=\tilde\xi F$. For $F\lesssim10^{-3}$ the whole range $\tilde\xi\le\tilde\xi_{\rm crit}$ satisfies both conditions for the three masses considered, and all the results of Sec.~\ref{sec:results} are therefore presented in terms of $\tilde\xi$, up to $\tilde\xi_{\rm crit}$.

\section{Parameter Constraints} \label{sec:results}

Having established the background physics of the injection term, here we want to present a systematic study on how the energy injection and the large cosmic birefringence angle generated from it is actually a model able to suppress the reionization bump in the case of a high value of the optical depth $\tau$. We have performed this analysis for axion masses $-31 \leq \log_{10}(m_\phi/\mathrm{eV}) \leq -30$, in steps of $0.1$, and for all the injection redshift windows shown in Fig.~\ref{fig:xicrit}, finding qualitatively similar results. In the following, we present the results for three representative mass values, $\log_{10}(m_\phi/\mathrm{eV}) = \lbrace -30.4, -30.5, -30.6\rbrace$, injecting energy only within the redshift window $50 \leq z \leq 60$.

\subsection{Data and methodology}
In this work we have considered the \textit{Planck} NPIPE low-$\ell$ \texttt{lollipop} data~\cite{Planck:2016mks, Tristram:2020wbi, Tristram:2021tvh}, considering the $EE$, $BB$, and $EB$ spectra and the corresponding NPIPE low-$\ell$ covariance matrix.
To compute the theoretical angular power spectra including the energy-injection mechanism described in Sec.~\ref{subsec:En_inj}, we have developed \texttt{birefINJ}\footnote{\url{https://github.com/SirlettiSS/birefINJ}}, a modified version of the Boltzmann code \texttt{CLASS}~\cite{CLASS, Blas:2011rf}.\ \texttt{birefINJ} implements the birefringence-induced source term of Eq.~\eqref{eq:Deltas} in the computation of the angular power spectra, Eq.~\eqref{Eq:ClXY}, following the approach of \texttt{birefCLASS}~\cite{Nakatsuka:2022}, together with the axion-field dynamics under energy injection described in Eq.~\eqref{eq:EoM_phi}.

Once the field dynamics and the resulting angular power spectra are computed, we evaluate the results through a $\chi^2$ statistics performed in a $EE$-only case or in a joint case involving $EE$, $BB$, and $EB$ spectra. We adopt as fiducial cosmology the best-fit $\Lambda$CDM model of Ref.~\cite{Sailer:2025:tau}, obtained without low-$\ell$ polarization data. When varying $\tau$, we keep $A_{\rm s}e^{-2\tau}$ fixed to its fiducial value, where $A_{\rm s}$ is the scalar amplitude, so that the high-$\ell$ CMB spectra are unchanged and only the low-$\ell$ reionization signal is affected. We do not include the $TT$ spectrum in the analysis. Cosmic birefringence does not affect the temperature anisotropies, so that $TT$ is independent of $\tilde\xi$; moreover, with $A_{\rm s}e^{-2\tau}$ fixed, increasing $\tau$ from $0.054$ to $0.09$ changes the low-$\ell$ $TT$ spectrum by less than $10\%$, well below the cosmic variance at these multipoles. We also do not include the low-$\ell$ $TE$ spectrum, whose constraining power on $\tau$ is limited by its large uncertainties.

Denoting by $\mathcal{S}$ the set of spectra entering a given analysis --- either $\mathcal{S} = \{EE\}$ for the EE-only case or $\mathcal{S} = \{EE, BB, EB\}$ for the joint analysis --- we define the data vector
\begin{equation}
\vec{D}^{\,X}(\tilde\xi,\tau) \equiv D_\ell^{X,\rm th}(\tilde\xi,\tau) - D_\ell^{X,\rm obs}, \qquad X \in \mathcal{S},
\end{equation}
and the concatenated vector $\vec{D}(\tilde\xi,\tau) = \bigoplus_{X \in \mathcal{S}} \vec{D}^{\,X}(\tilde\xi,\tau)$, obtained by stacking each spectrum in $\mathcal{S}$. Here $D^{X}_\ell = \frac{\ell(\ell+1)}{2\pi} C_\ell^{X}$ is the angular power spectrum in the standard $D_\ell$ convention. The $\chi^2$ is then
\begin{equation}
\chi^2(\tilde\xi,\tau) = \vec{D}(\tilde\xi,\tau)^{T} \cdot \, \text{Cov}^{-1} \cdot  \, \vec{D}(\tilde\xi,\tau) \, ,
\end{equation}
where $\text{Cov}$ is the NPIPE low-$\ell$ covariance matrix restricted to the spectra in $\mathcal{S}$: for the EE-only case $\text{Cov}$ reduces to the single $EE$ block, while for the joint analysis $\text{Cov}$ is the full $3N_\ell \times 3N_\ell$ block covariance correlating $EE$, $BB$, and $EB$, with $N_\ell$ the number of multipoles. In this analysis $N_\ell = 29$, having restricted the analysis to the multipole range $2 \leq \ell \leq 30$. The corresponding likelihood is $\mathcal{L}(\tilde\xi,\tau) \propto e^{-\chi^2(\tilde\xi,\tau)/2}$.
We evaluate $\chi^2(\tilde\xi,\tau)$ using two independent methods: a Markov Chain Monte Carlo (MCMC) sampler and a direct evaluation on a two-dimensional $(\tilde\xi,\tau)$ grid at fixed ALP mass. The two approaches have been cross-validated against each other, yielding consistent results; in the following we present the results obtained from the grid-based likelihood evaluation.

\subsection{Power spectra}\label{sec:power_spectra}

\begin{figure*}[t]
  \centering
  \begin{minipage}[t]{0.48\textwidth}
    \centering
    \includegraphics[width=\linewidth]{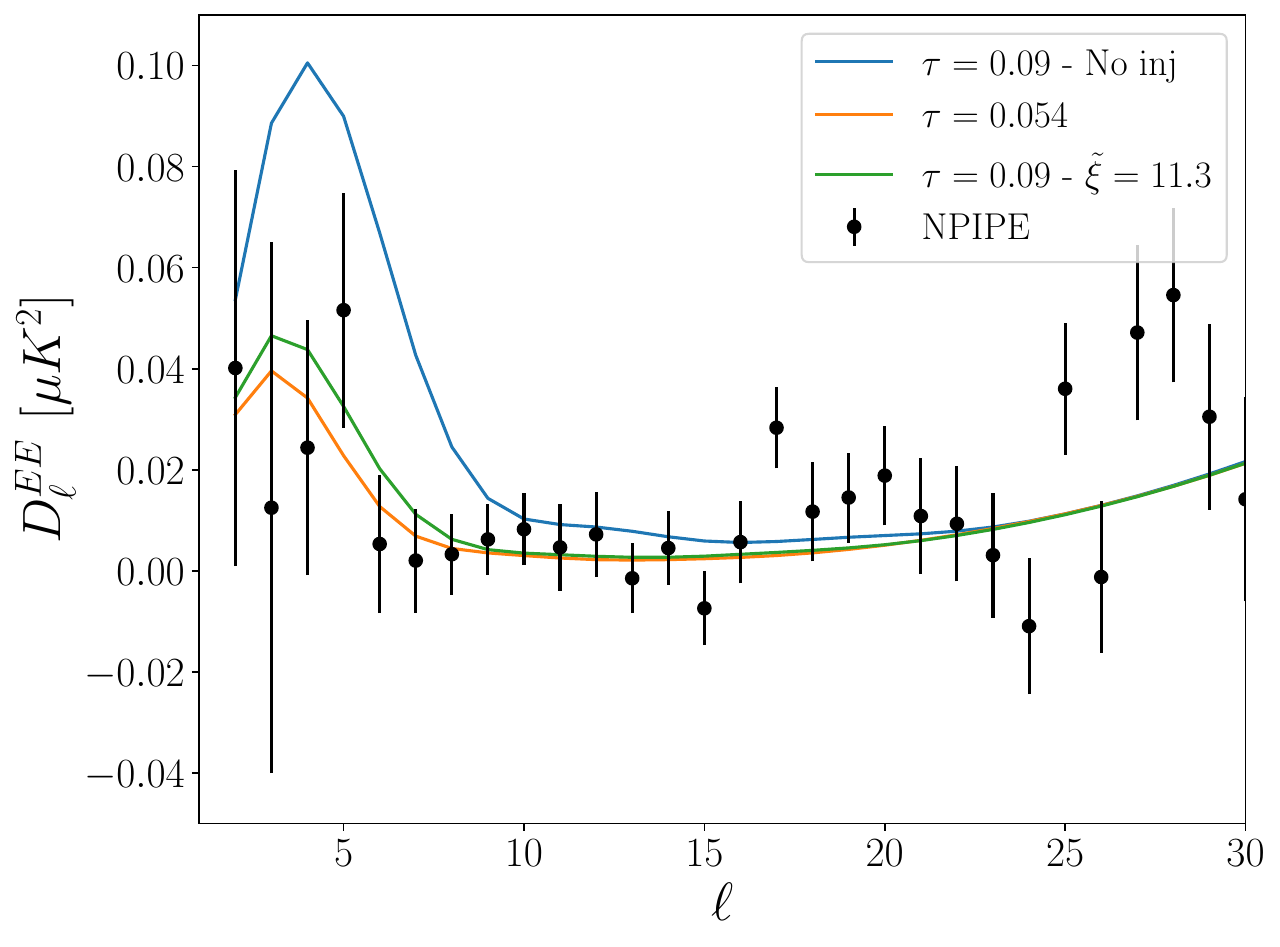}
  \end{minipage}\hfill
  \begin{minipage}[t]{0.48\textwidth}
    \centering
    \includegraphics[width=\linewidth]{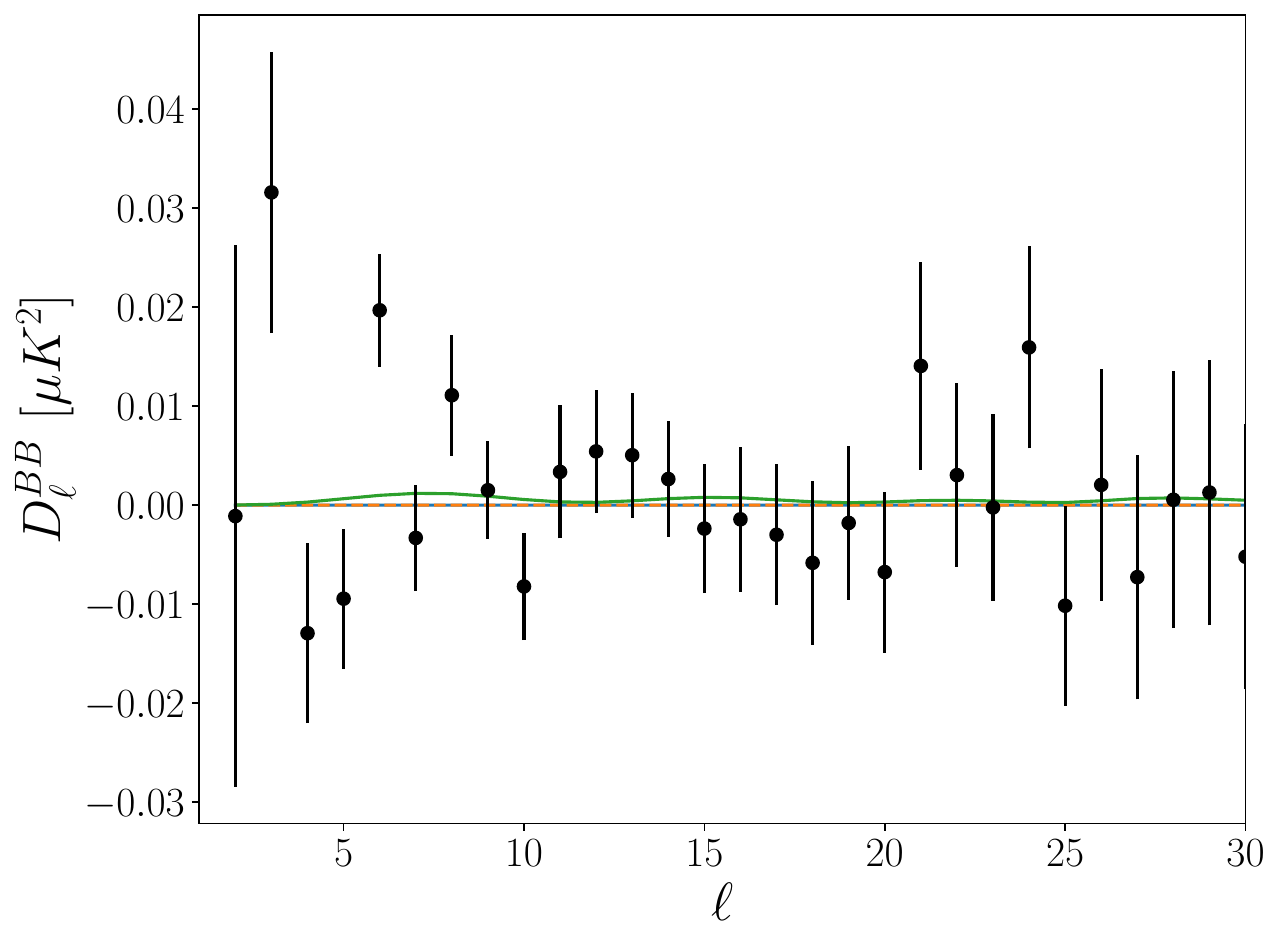}
  \end{minipage}

  \vspace{0.5em}

  \begin{minipage}[t]{0.48\textwidth}
    \centering
    \includegraphics[width=\linewidth]{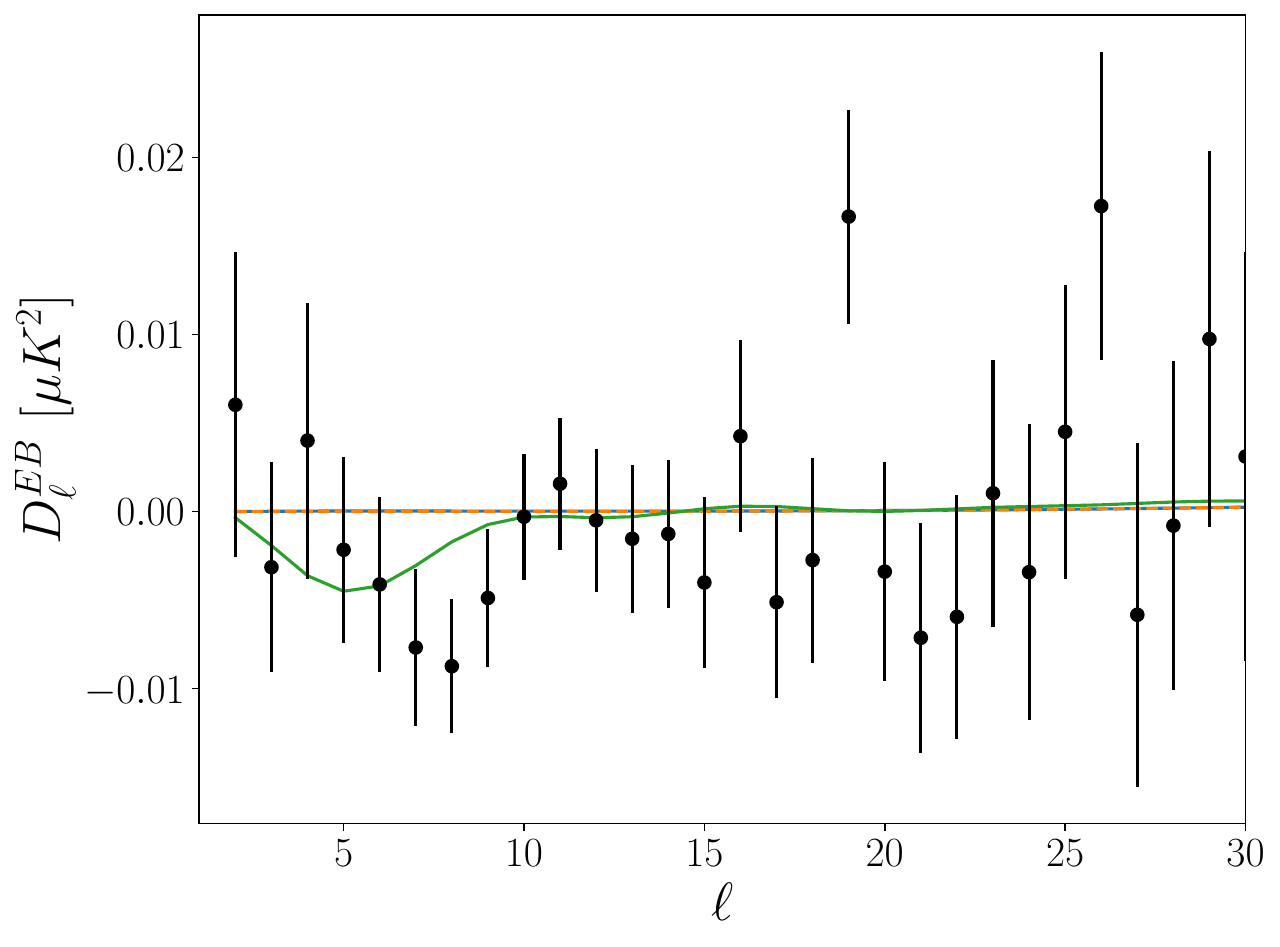}
  \end{minipage}

  \caption{$EE$ (upper left), $BB$ (upper right), and $EB$ (center) angular power spectra for $\log_{10}(m_\phi/\mathrm{eV}) = -30.5$, compared to the \textit{Planck} NPIPE low-$\ell$ data (black points). Orange: $\tau = \tau_{\rm Planck}=0.054$ with no injection. Blue: $\tau = 0.09 >\tau_{\rm Planck} $ with no injection where the resulting $EE$ bump overshoots the observed reionization signal. Green: $\tau = 0.09$ but with an injection of $\tilde\xi = 11.3$, below the kick-out limit for this mass, $\tilde\xi_{\rm crit}= 16.3$; in this case, the injection suppresses the $EE$ bump back to a level consistent with the data. The same injection also imprints a distinctive dip in $D_\ell^{EB}$ and a corresponding bump in $D_\ell^{BB}$ around $\ell \sim 5$--$8$, the direct signature of the late-time rotation sourced by the axion dynamics.}
  \label{fig:spectra}
\end{figure*}

As recently proposed in the literature~\cite{Giare:2023:tau,Loverde:2024:mnu,Sailer:2025:tau,Jhaveri:2025:tau}, raising the optical depth to $\tau \sim 0.09$ can help alleviate the CMB-BAO tension. However, at this value of $\tau$ and in the absence of injection, the resulting $EE$ reionization bump clearly overshoots the \textit{Planck} data (blue curve in Fig.~\ref{fig:spectra}). Introducing a large cosmic birefringence angle sourced by the energy-injection mechanism of Eq.~\eqref{eq:EoM_phi} suppresses the bump back down to a level consistent with the data (green curve in Fig.~\ref{fig:spectra}), closely tracking the standard, low-$\tau$ case (orange curve in Fig.~\ref{fig:spectra}). The same qualitative behavior, but with different values of $\tilde\xi$ and $\tau$, is observed for the other masses considered in this work and is therefore not shown separately.

The $BB$ and $EB$ panels of Fig.~\ref{fig:spectra} show that the injection produces a genuine, visible change in these two channels as well. However, given the current NPIPE low-$\ell$ uncertainties, all three curves remain statistically consistent with the data in both channels: no single spectrum, taken individually, visually excludes either scenario. This suggests that $EB$ and $BB$ can carry complementary information to that of $EE$ alone, which tightly constrains the shape of the reionization bump but leaves $(\tilde\xi,\tau)$ highly degenerate. As shown quantitatively in Sec.~\ref{sec:contours}, combining $EE$, $BB$, and $EB$ reshapes the allowed $(\tilde\xi,\tau)$ region into a set of distinct sub-regions rather than a single, simply-connected area; a residual degeneracy nonetheless persists, as we discuss in Sec.~\ref{sec:contours}.

Altogether, Fig.~\ref{fig:spectra} shows that there exist values of $(\tilde\xi,\tau)$ for which a large optical depth remains compatible with the observed $EE$, $BB$, and $EB$ spectra, thereby strongly motivating the quantitative parameter analysis presented in the remainder of this section. We note that all these results closely resemble those obtained by Ref.~\cite{Namikawa:2025:PRL} within the $n\pi$-phase model.

\subsection{Contour plots}\label{sec:contours}

Having shown in Sec.~\ref{sec:power_spectra} that a large optical depth can, for suitable values of $(\tilde\xi,\tau)$, remain qualitatively compatible with the observed $EE$, $BB$, and $EB$ spectra, we now turn to a quantitative likelihood analysis of the full $(\tilde\xi,\tau)$ parameter space. We compare the results with the $EE$-only and joint $EE$+$BB$+$EB$ cases under both a uniform and a Gaussian prior on $\tau$.

\begin{figure*}[!t]
\centering

\includegraphics[width=\textwidth,height=0.46\textheight,keepaspectratio]{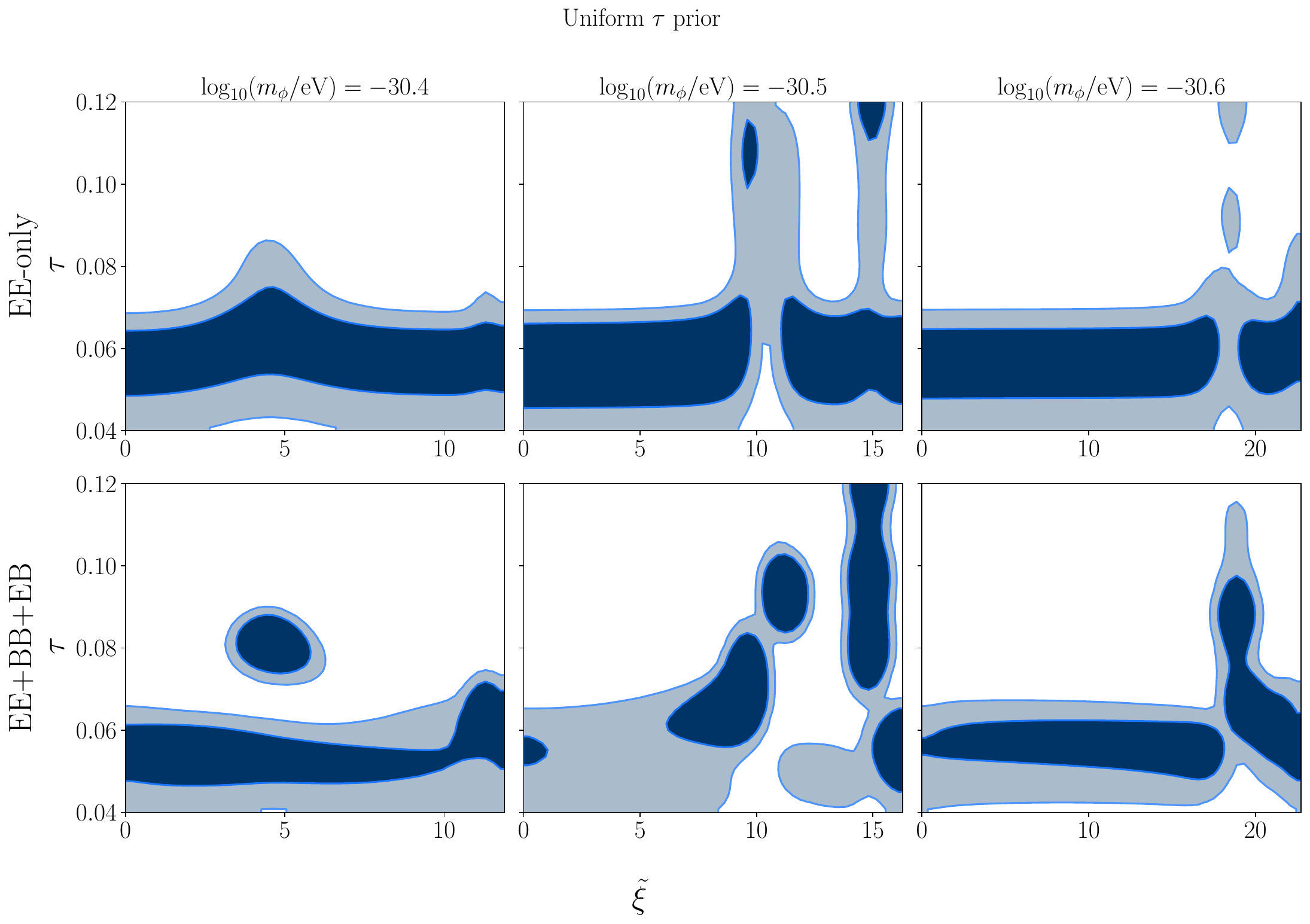}

\vspace{0.3em}

\includegraphics[width=\textwidth,height=0.46\textheight,keepaspectratio]{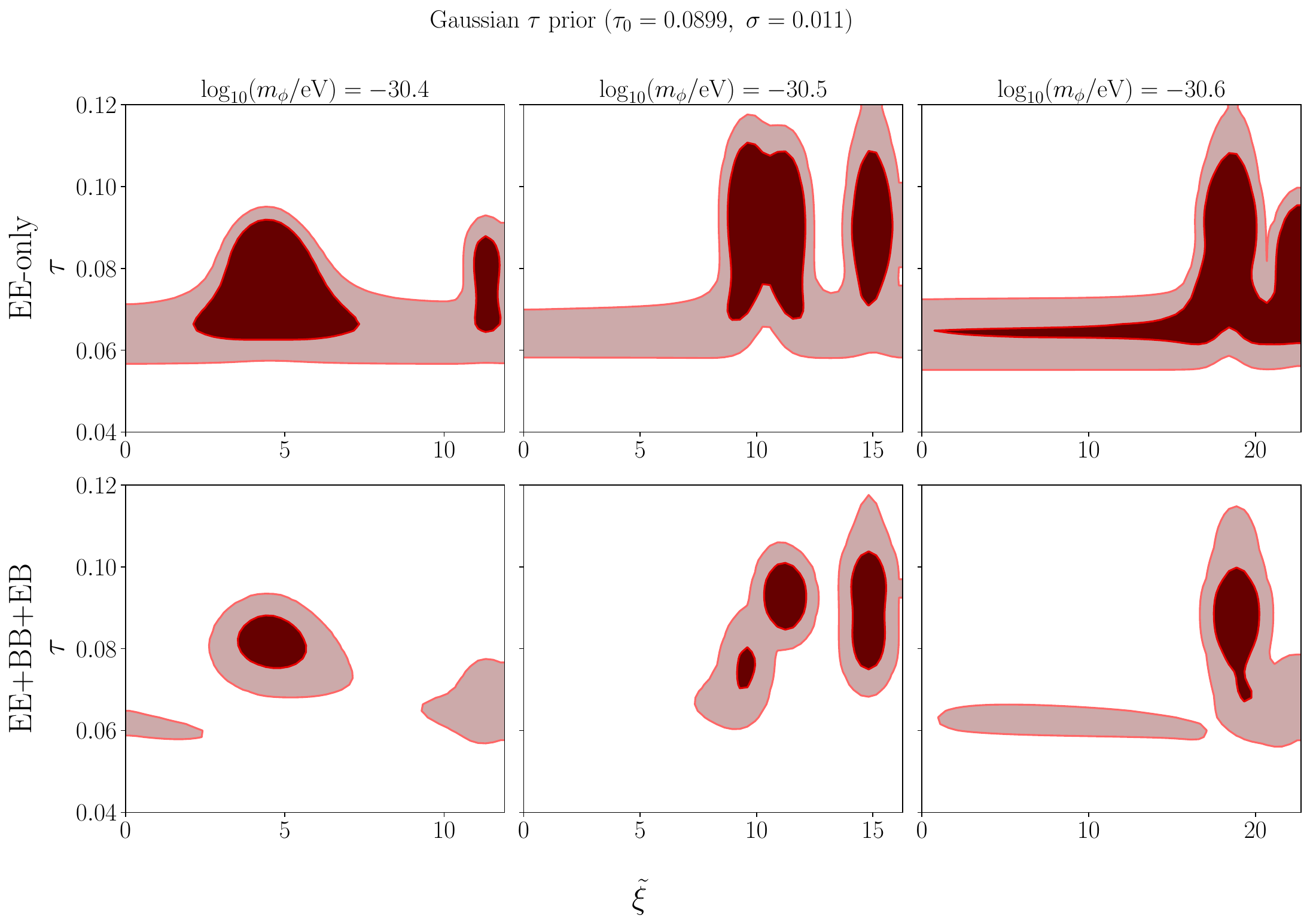}

\caption{$1\sigma$ (dark) and $2\sigma$ (light) confidence contours in the $(\tilde\xi,\tau)$ plane for the $EE$-only (top row of each block) and joint $EE$+$BB$+$EB$ (bottom row of each block) analyses, for the three ALP masses $\log_{10}(m_\phi/\mathrm{eV}) = -30.4, -30.5, -30.6$ (columns). \emph{Top block:} uniform prior $\tau\in(0.04,0.12)$. \emph{Bottom block:} Gaussian prior $\tau\sim\mathcal{N}(\tau_0=0.0899,\sigma=0.011)$, motivated by high-$\ell$ CMB+DESI-BAO constraints~\cite{Sailer:2025:tau}. Note that the $\tilde\xi$ range differs across columns, as $\tilde\xi_{\rm crit}$ depends on the ALP mass (cf. Fig.~\ref{fig:xicrit}).}
\label{fig:contours}
\end{figure*}

Top block of Fig.~\ref{fig:contours} shows the $1\sigma$ and $2\sigma$ confidence contours in the $(\tilde\xi,\tau)$ plane obtained assuming a uniform prior on $\tau$. In the $EE$-only case (top row), all three masses display a common qualitative structure: a broad, continuous region around $\tau \sim 0.05$--$0.07$, together with one or more narrow features extending toward higher $\tau$ at specific, discrete values of $\tilde\xi$. This multimodal structure is not a numerical artifact --- it reflects the oscillatory nature of the ALP dynamics under a finite-window energy injection, whereby the accumulated rotation phase $\beta(\eta)$ can reach similar values for several distinct values of $\tilde\xi$, producing comparably good fits to the $EE$ reionization bump at those discrete points.

Combining $EE$, $BB$, and $EB$ (bottom row) does not simply narrow this structure into a single, better-constrained region. Instead, for all three masses the allowed parameter space is reorganized into a set of distinct, sometimes disconnected sub-regions, some of which persist at values of $\tau$ higher than the dominant low-$\tau$ feature of the $EE$-only case. This is consistent with the fact that $EE$, $BB$, and $EB$ are not independent probes: all three spectra are sourced by the same underlying rotation history where $\beta=\beta(\eta,\tilde\xi,\tau)$, so the additional information carried by $BB$ and $EB$ reshapes rather than simply shrinks the $EE$-only degeneracy.

Bottom block of Fig.~\ref{fig:contours} shows the same analysis with a Gaussian prior on $\tau$ centered at $\tau_0 = 0.0899$, motivated by external CMB+BAO constraints. As expected, for all three masses, this prior consistently selects and reinforces the higher-$\tau$ sub-regions that were only marginal features under a uniform prior, while suppressing the low-$\tau$ region that dominates the $EE$-only, uniform-prior case. In all three cases, the surviving sub-regions cluster around $\tau \sim 0.08$--$0.10$, directly tracking the external prior value $\tau_0 = 0.0899$. This result is robust across the three representative masses considered in this work: there consistently exist regions of the $(\tilde\xi,\tau)$ parameter space, compatible with the joint $EE$+$BB$+$EB$ likelihood, in which a large optical depth is achieved through the energy-injection mechanism.

\begin{figure*}[t]
\centering
\includegraphics[width=\textwidth]{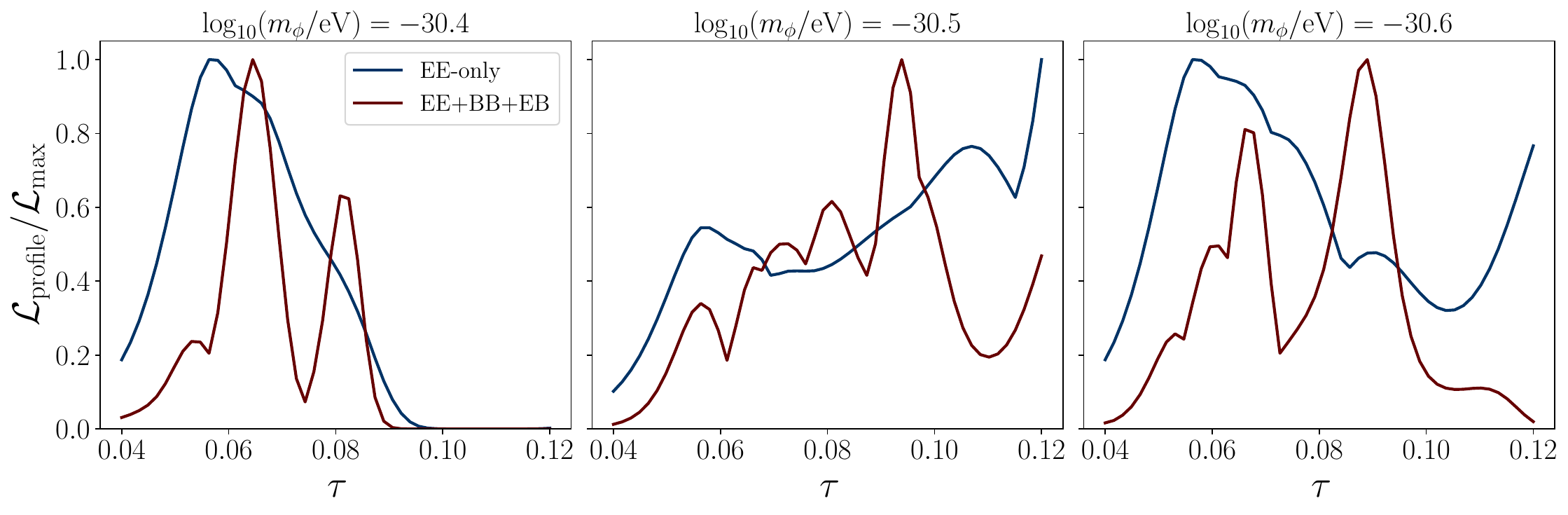}
\caption{Profile likelihood in $\tau$, $\mathcal{L}_{\rm profile}(\tau)=\max_{\tilde\xi}\mathcal{L}(\tilde\xi,\tau)$, normalized to its maximum, for the $EE$-only (blue) and joint $EE$+$BB$+$EB$ (red) analyses and the three ALP masses of Fig.~\ref{fig:contours}.}
\label{fig:profile-tau}
\end{figure*}

Because the allowed parameter space is multimodal rather than unimodal, a single symmetric credible interval does not adequately summarize the joint constraint; we therefore refrain from quoting a single best-fit value with symmetric errors, and instead rely on the sub-regions shown in the two blocks of Fig.~\ref{fig:contours} to convey the structure of the allowed parameter space. To further characterize this multimodal structure, Fig.~\ref{fig:profile-tau} shows the profile likelihood in $\tau$, obtained by maximizing the likelihood over $\tilde\xi$ at each value of $\tau$. Unlike marginalization, this procedure is independent of the prior volume in $\tilde\xi$ and is therefore not affected by volume effects in the degenerate regime. In the joint analysis the profile exhibits several distinct peaks, including some at $\tau\gtrsim 0.08$, consistent with the large-$\tau$ regions of Fig.~\ref{fig:contours} being genuine features of the likelihood.
\\

\section{Summary and Discussion} \label{sec:discussion}
We quantitatively explored the energy injection scenario in which a tiny fraction of the CDM energy density is transferred to the ALP energy density before the reionization epoch. Using a specific model of energy injection, we numerically showed that the ALP field value is enhanced after the energy injection. We then showed that the $EE$ power spectrum at low-$\ell$ is suppressed due to a large rotation angle of cosmic birefringence after the reionization epoch. We found a parameter region of the model that simultaneously explains the \textit{Planck} low-$\ell$ polarization data, a nonzero value of the rotation angle of cosmic birefringence obtained from $EB$ power spectrum at high-$\ell$, and a large value of $\tau$ suggested as a possible solution to the CMB-BAO tension. Although we have shown results for a single injection window and three representative masses, we have verified that similar parameter regions exist for all ALP masses $-31 \leq \log_{10}(m_\phi/\mathrm{eV}) \leq -30$ and injection windows within $10 \leq z \leq 100$.

In our constraints on the model parameters, we did not include the instrumental miscalibration angle. The miscalibration angle could introduce a nonzero $EB$ power spectrum. However, both the $EE$-only and $EE+EB+BB$ cases allow a larger value of $\tau$ at a similar value of $\tilde{\xi}$, suggesting that the impact of the miscalibration angle is not significant. The miscalibration angles at each frequency channel constrained from Ref.~\cite{Eskilt:2022:biref-const} are also within sub-degree, indicating that the impact of the miscalibration angle is much smaller than the observational errors of the $EB$ power spectra at low-$\ell$. 

Our model predicts a large rotation angle of cosmic birefringence at low redshift, $z<10$. Such a large rotation angle can be tested by observations of radio galaxies \cite{Carroll:1997:radio,Naokawa:2025shr}. Measurements of the low-$\ell$ polarization by future CMB experiments, such as LiteBIRD \cite{LiteBIRD}, are also important to validate the measurement of low-$\ell$ polarization power spectra.


\begin{acknowledgements}
We thank Elisa G. M. Ferreira for valuable suggestions at the early stage of this project. We also thank Luca Visinelli for useful comments on the draft. We acknowledge support from JSPS KAKENHI Grant No. JP22K03682, No. JP24KK0248, and No. JP25K00996, and from JST EXPERT-J, Japan Grant No. JPMJEX2508. The Kavli IPMU is supported by World Premier International Research Center Initiative (WPI Initiative), MEXT, Japan. This work made use of the \texttt{idark} computing cluster at the Kavli IPMU.  
\end{acknowledgements}


\section*{Data availability}
The data that support the findings of this article are publicly available at the following GitHub repository: \url{https://github.com/SirlettiSS/birefINJ}.

\onecolumngrid

\appendix

\section{Energy injection and the spectator condition}\label{app:phi_thermodynamics}

In this appendix, we derive the evolution of the CDM and ALP energy densities during and after the injection, which are the basis for the depletion and spectator conditions discussed in Sec.~\ref{sec:constraints_xiF}. We denote by $a_{\rm start}=1/(1+z_{\rm start})$ and $a_{\rm end}=1/(1+z_{\rm end})$ the scale factors at the beginning and at the end of the injection window, and by $a_{\rm ini}\ll a_{\rm start}$ the scale factor at which the initial conditions are set.

\subsection{Dark matter evolution and depletion limit}

The CDM density obeys Eq.~\eqref{eq:densityeqCDM} with the source $Q$ of Eq.~\eqref{eq:Q}, i.e.
\begin{equation}
    \bar\rho'_{\rm dm}+3\mathcal{H}\bar\rho_{\rm dm}=-Q,\qquad
    Q= \Theta(a) \xi \mathcal{H} \bar\rho_{\rm dm} =\begin{cases}\xi \mathcal{H} \bar\rho_{\rm dm} & a_{\rm start}\le a\le a_{\rm end},\\ 0 & \text{otherwise,}\end{cases}
\end{equation}
where $\Theta(a)$ is the step-function, equal to $1$ inside the injection window and $0$ outside.
Since $Q\propto \mathcal{H}$, the Hubble rate cancels when the equation is written in terms of the scale factor,
\begin{equation}
    \frac{d\bar\rho_{\rm dm}}{da}=-\frac{3+\xi\,\Theta(a)}{a}\,\bar\rho_{\rm dm}.
\end{equation}
The equation can then be solved separately in the three intervals, requiring continuity of $\bar\rho_{\rm dm}$ at $a_{\rm start}$ and $a_{\rm end}$. Before the injection, there is no source, and $\bar\rho_{\rm dm}$ coincides with the $\Lambda$CDM solution $\bar\rho_{\rm dm}^{\Lambda{\rm CDM}}(a)=\bar\rho_{\rm dm,0}\,a^{-3}$, where $\bar\rho_{\rm dm,0}$ is the present-day $\Lambda$CDM value. We obtain 
\begin{equation}\label{eq:rhodm_phases}
    \bar\rho_{\rm dm}(a)=\bar\rho_{\rm dm,0}\,a^{-3}\times
    \begin{cases}
        1, & a\le a_{\rm start},\\[4pt]
        \left(\dfrac{a}{a_{\rm start}}\right)^{-\xi}, & a_{\rm start}\le a\le a_{\rm end},\\[10pt]
        \left(\dfrac{a_{\rm end}}{a_{\rm start}}\right)^{-\xi}, & a\ge a_{\rm end}.
    \end{cases}
\end{equation}
In the window, the density therefore scales as $a^{-(3+\xi)}$, while after the injection it dilutes again as $a^{-3}$, but with a reduced normalization.

Dividing by the $\Lambda$CDM density, the depletion factor after the injection is (cf. Eq.\eqref{eq:deplationlimit})
\begin{equation}
    R(\xi)\equiv\frac{\bar\rho_{\rm dm}}{\bar\rho_{\rm dm}^{\Lambda{\rm CDM}}}\bigg|_{a\ge a_{\rm end}}
    =\left(\frac{a_{\rm end}}{a_{\rm start}}\right)^{-\xi}
    =\left(\frac{1+z_{\rm start}}{1+z_{\rm end}}\right)^{-\xi},
\end{equation}
which depends only on $\xi$ and on the width of the injection window, and is independent of $m_\phi$ and $F$. Imposing the upper bound from the data as discussed in Sec.~\ref{sec:constraints_xiF}, $1-R(\xi)<10^{-2}$, i.e.\ $R(\xi)>0.99$, gives Eq.~\eqref{eq:xi_dep}. For the window $50\le z\le60$ used throughout the paper, $\xi_{\rm dep}\simeq0.056$.

\subsection{ALP energy density}

The energy density of the ALP obeys the continuity equation with the source $Q$ of Eq.~\eqref{eq:Q};
\begin{equation}
    \bar\rho'_\phi+3\mathcal{H}(1+w_\phi)\bar\rho_\phi=\xi \mathcal{H}\bar\rho_{\rm dm},
\end{equation}
with $w_\phi=\bar{p}_\phi/\bar{\rho}_\phi$. Using the scale factor as a time variable, this reads
\begin{equation}\label{eq:rhophi_a}
    a\,\frac{d\bar\rho_\phi}{da}+3(1+w_\phi)\,\bar\rho_\phi=\xi\,\bar\rho_{\rm dm}(a),
\end{equation}
with $\bar\rho_{\rm dm}(a)$ given by Eq.~\eqref{eq:rhodm_phases}, and the initial condition $\bar\rho_\phi(a_{\rm ini})=\bar\rho_{\phi,\rm ini}=F\bar\rho_{\rm dm,0}$. Assuming $w_\phi$ constant, the equation can be solved in the three intervals.
\\

\paragraph{Interval 1, $a\le a_{\rm start}$.} There is no source and
\begin{equation}
    \bar\rho_\phi(a)=\bar\rho_\phi(a_{\rm ini})\left(\frac{a}{a_{\rm ini}}\right)^{-3(1+w_\phi)}.
\end{equation}

\paragraph{Interval 2, $a_{\rm start}< a\le a_{\rm end}$.} The equation to solve is
\begin{equation}
    \frac{d\bar\rho_\phi}{da}+\frac{3(1+w_\phi)}{a}\bar\rho_\phi=\frac{\xi}{a}\,\bar\rho_{\rm dm}(a_{\rm ini})\left(\frac{a_{\rm start}}{a_{\rm ini}}\right)^{-3}\left(\frac{a}{a_{\rm start}}\right)^{-(3+\xi)}.
\end{equation}
It is linear and can be solved exactly with the integrating factor $a^{3(1+w_\phi)}$, which turns the left-hand side into $a^{-3(1+w_\phi)}\,d\left[a^{3(1+w_\phi)}\bar\rho_\phi\right]/da$. The solution is
\begin{equation}
    \bar\rho_\phi(a)=\bar\rho_\phi(a_{\rm start})\left(\frac{a}{a_{\rm start}}\right)^{-3(1+w_\phi)}
    +\frac{\xi}{3w_\phi-\xi}\,\bar\rho_{\rm dm}(a_{\rm start})
    \left[\left(\frac{a}{a_{\rm start}}\right)^{-(3+\xi)}-\left(\frac{a}{a_{\rm start}}\right)^{-3(1+w_\phi)}\right],
\end{equation}
with the boundary values
\begin{equation}
    \bar\rho_\phi(a_{\rm start})=\bar\rho_\phi(a_{\rm ini})\left(\frac{a_{\rm start}}{a_{\rm ini}}\right)^{-3(1+w_\phi)},\qquad
    \bar\rho_{\rm dm}(a_{\rm start})=\bar\rho_{\rm dm}(a_{\rm ini})\left(\frac{a_{\rm start}}{a_{\rm ini}}\right)^{-3}.
\end{equation}
At $a=a_{\rm end}$ this gives
\begin{equation}\label{eq:constant_term}
\begin{split}
    \bar\rho_\phi(a_{\rm end})={}&\bar\rho_\phi(a_{\rm ini})\left(\frac{a_{\rm end}}{a_{\rm ini}}\right)^{-3(1+w_\phi)}\\
    &+\bar\rho_{\rm dm}(a_{\rm ini})\left(\frac{a_{\rm start}}{a_{\rm ini}}\right)^{-3}\frac{\xi}{3w_\phi-\xi}
    \left[\left(\frac{a_{\rm end}}{a_{\rm start}}\right)^{-(3+\xi)}-\left(\frac{a_{\rm end}}{a_{\rm start}}\right)^{-3(1+w_\phi)}\right].
\end{split}
\end{equation}

\paragraph{Interval 3, $a>a_{\rm end}$.} The source vanishes again, and
\begin{equation}\label{eq:rhophi_phase3}
    \bar\rho_\phi(a)=\bar\rho_\phi(a_{\rm end})\left(\frac{a}{a_{\rm end}}\right)^{-3(1+w_\phi)}.
\end{equation}

\subsection{Spectator ratio and the function $\mathcal{G}$}

The ratio between the ALP and CDM densities follows from writing $\bar\rho_{\rm dm}(a)=\bar\rho_{\rm dm,0}\,a^{-3}R(a)$ and $\bar\rho_{\phi,\rm ini}=F\bar\rho_{\rm dm,0}$,
\begin{equation}\label{eq:ratiospec}
    \frac{\bar\rho_\phi(a)}{\bar\rho_{\rm dm}(a)}=F\,\frac{1}{R(a)}\,a^3\,\frac{\bar\rho_\phi(a)}{\bar\rho_{\phi,\rm ini}},
\end{equation}
which for $a=1$ is Eq.~\eqref{eq:ratio_a1}. The dependence of $\mathcal{G}\equiv[a^3\bar\rho_\phi/\bar\rho_{\phi,\rm ini}]_{a=1}$, defined in Eq.~\eqref{eq:mathcalG}, on $F$ and $\xi$ follows from the solution in the third interval, Eq.~\eqref{eq:rhophi_phase3}, once the constant term of Eq.~\eqref{eq:constant_term} is expressed in terms of $F$ and $\xi$. Evaluating it at $a=1$ and using $\tilde\xi=\xi/F$ we obtain
\begin{equation}\label{eq:G_explicit}
    \mathcal{G}=\Bigg\lbrace\left(\frac{a_{\rm end}}{a_{\rm ini}}\right)^{-3(1+w_\phi)}
    +\tilde\xi\,\frac{a_{\rm start}^{-3}}{3w_\phi-\xi}
    \left[\left(\frac{a_{\rm end}}{a_{\rm start}}\right)^{-(3+\xi)}-\left(\frac{a_{\rm end}}{a_{\rm start}}\right)^{-3(1+w_\phi)}\right]\Bigg\rbrace\,a_{\rm end}^{3(1+w_\phi)},
\end{equation}
where for the second term we have used $\bar\rho_{\rm dm}(a_{\rm ini})=\bar\rho_{\rm dm,0}\,a_{\rm ini}^{-3}$ (there is no depletion before $a_{\rm start}$) and $\bar\rho_\phi(a_{\rm ini})=\rho_{\phi,\rm ini}=F\bar\rho_{\rm dm,0}$ to obtain
\begin{equation}
    \frac{\bar\rho_{\rm dm}(a_{\rm ini})}{\bar\rho_\phi(a_{\rm ini})}\left(\frac{a_{\rm start}}{a_{\rm ini}}\right)^{-3}=\frac{a_{\rm start}^{-3}}{F}.
\end{equation}

The first term in Eq.~\eqref{eq:G_explicit} is the free evolution of the initial energy and does not depend on $F$ or $\xi$, while the injected energy is proportional to $\tilde\xi$. Therefore, $\mathcal{G}$ depends on $F$ and $\xi$ only through $\tilde\xi=\xi/F$, up to the explicit dependence on $\xi$ in $3w_\phi-\xi$ and $(a_{\rm end}/a_{\rm start})^{-\xi}$, which is at the percent level for $\xi\le\xi_{\rm dep}$. The ratio at $a=1$ is then $F\,\mathcal{G}(\tilde\xi)/R(\xi)$, which is the expression used in Eq.~\eqref{eq:ratio_a1} to compute the ALP spectator limit in Eq.~\eqref{eq:spectatorlimit}.

\subsection{Limits of the analytical approximation}

\begin{figure}[!t]
    \centering
    \includegraphics[width=\textwidth]{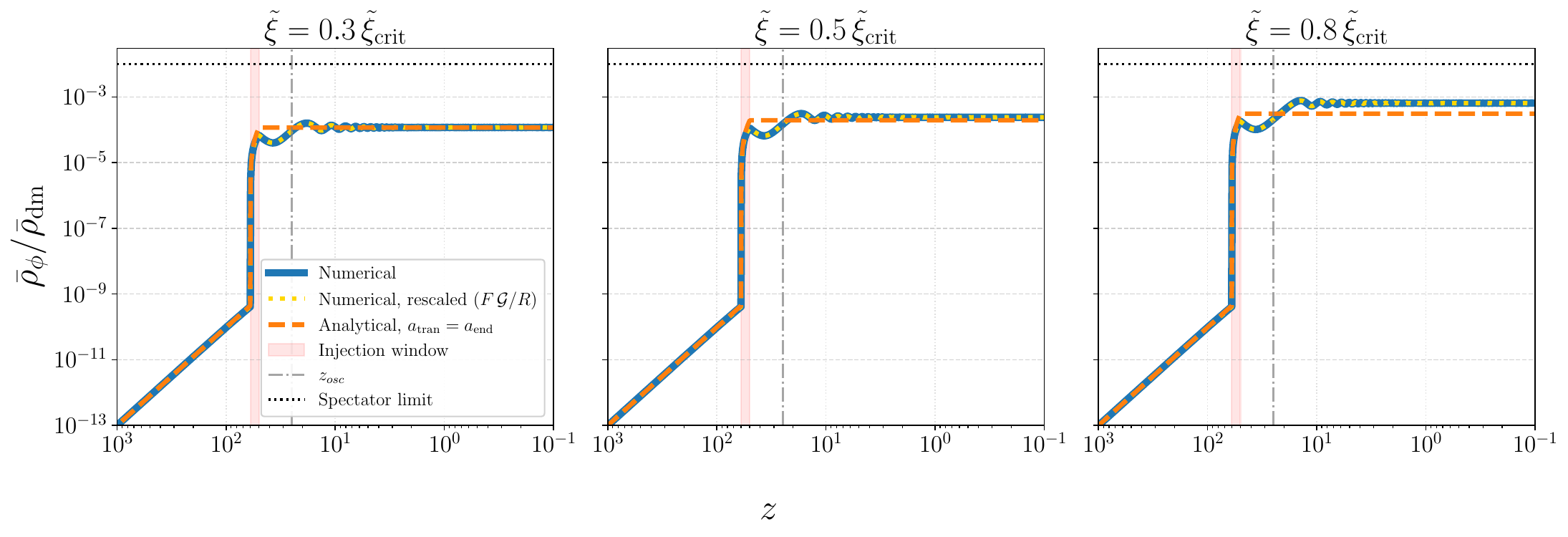}
    \caption{Evolution of the ratio $\bar\rho_\phi/\bar\rho_{\rm dm}$ for $\log_{10}(m_\phi/{\rm eV})=-30.5$ and $F=10^{-4}$, for $\tilde\xi=0.3$ (left), $0.5$ (center) and $0.8\,\tilde\xi_{\rm crit}$ (right). The shaded band marks the injection window $50\le z\le60$, the dash-dotted line marks $z_{\rm osc}\simeq25$, and the dotted horizontal line the spectator limit, Eq.~\eqref{eq:spectatorlimit}. The thick blue line is the direct numerical solution of Eq.~\eqref{Eq:x-EoM:inject} at the given $(\xi,F)$. The dotted yellow line is obtained by solving the equation once at fixed $\tilde\xi$ and rescaling the result as $F\,\mathcal{G}/R$, as done to compute the contours of Fig.~\ref{fig:exclusion_plot_logm-30.5}. Its agreement with the direct numerical solution (thick blue) validates this construction. The dashed orange line is the analytical solution with an instantaneous transition from $w_\phi=-1$ to $w_\phi=0$ at $a_{\rm end}$. It reproduces the numerical result well for small $\tilde\xi$, but progressively underestimates it as $\tilde\xi$ increases. This is due to the intermediate phase between $a_{\rm end}$ and $a_{\rm osc}$, in which $w_\phi=w_\phi(a)$ varies and is not captured by the sharp transition: its duration grows with $\tilde\xi$, and for $\tilde\xi$ approaching $\tilde\xi_{\rm crit}$ it extends beyond $a_{\rm osc}$, as shown by the oscillations of the numerical curve, which are larger and last longer in the right panel.}
    \label{fig:ratio_analytic_vs_numeric}
\end{figure}

Equation~\eqref{eq:G_explicit} is obtained for constant $w_\phi$. To first approximation, the evolution of the field can be described by a sharp and instantaneous transition, at a scale factor $a_{\rm tran}$, from $w_\phi=-1$, when the field is frozen in the potential by Hubble friction, to $w_\phi=0$, when it oscillates around the minimum and behaves as matter. A natural choice is to place the transition at the end of the injection, $a_{\rm tran}=a_{\rm end}$. For $w_\phi=-1$ up to $a_{\rm tran}$ and $w_\phi=0$ afterwards, Eq.~\eqref{eq:G_explicit} becomes
\begin{equation}
    \mathcal{G}=a_{\rm tran}^3\left\{1+\tilde\xi\,\frac{a_{\rm start}^{-3}}{3+\xi}\left[1-\left(\frac{a_{\rm end}}{a_{\rm start}}\right)^{-(3+\xi)}\right]\right\}.
\end{equation}
This is a good approximation of the exact result. As an example, Fig.~\ref{fig:ratio_analytic_vs_numeric} compares the analytical ratio $\bar\rho_\phi/\bar\rho_{\rm dm}$ with the transition at $a_{\rm end}$ to the numerical solution.

In reality, the transition from $w_\phi=-1$ to $w_\phi=0$ is smooth. After the end of the injection, the field is still frozen in the potential until the Hubble friction becomes comparable to the mass, $3H(a_{\rm osc})\simeq m_\phi$. In the matter era, $H(a)\simeq H_0\sqrt{\Omega_{\rm m}}\,a^{-3/2}$, which gives
\begin{equation}
    a_{\rm osc}=\left(\frac{3H_0\sqrt{\Omega_{\rm m}}}{m_\phi}\right)^{2/3}.
\end{equation}
Between $a_{\rm end}$ and $a_{\rm osc}$, the equation of state is not constant but varies with redshift $w_\phi=w_\phi(a)$. This transition phase can end or extend even beyond $a_{\rm osc}$, depending on how close $\tilde\xi$ is to its critical value. A strong injection gives the field a large velocity, such that its kinetic energy $\tfrac12\dot\phi^2$ dominates over the potential energy, and $w_\phi$ can reach $+1$ (kination). In this regime, the kinetic energy dilutes as $a^{-6}$, faster than matter, so that the ALP energy density decreases quickly until the field is again frozen by Hubble friction ($w_\phi\to-1$), and the onset of the oscillations is delayed. Since this behavior is not captured by a constant $w_\phi$, the sharp-transition approximation can both overestimate and underestimate the ratio, depending on $\tilde\xi$.

Although the analytical approximation works well in most of the parameter space, we generate the exclusion plot shown in Fig.~\ref{fig:exclusion_plot_logm-30.5} from the numerical solution of Eq.~\eqref{Eq:x-EoM:inject}, which treats the evolution of $w_\phi$ exactly for every $\tilde\xi$, following the procedure described in Sec.~\ref{sec:constraints_xiF}.

\twocolumngrid

\bibliographystyle{mybst}
\bibliography{cite}

\end{document}